\documentclass[conference]{IEEEtran}
\ifCLASSINFOpdf
   \usepackage[pdftex]{graphicx}
   \graphicspath{{../pdf/}{../jpeg/}}
   \DeclareGraphicsExtensions{.pdf,.jpeg,.png}
\else
   \usepackage[dvips]{graphicx}
   \graphicspath{{../eps/}}
   \DeclareGraphicsExtensions{.eps}
\fi
\begin{document}
%
\title{Exact calculation of current correlations and admittance in the fractional quantum Hall regime}

\author{\IEEEauthorblockN{Redouane Zamoum}
\IEEEauthorblockA{Aix-Marseille Universit\'e,
Universit\'e de Toulon\\
CNRS, CPT UMR 7332,
13288, Marseille, France\\
Email: zamoum.redouane@gmail.com}
\and
\IEEEauthorblockN{Adeline Cr\'epieux}
\IEEEauthorblockA{Aix-Marseille Universit\'e,
Universit\'e de Toulon\\
CNRS, CPT UMR 7332,
13288, Marseille, France\\
Email: adeline.crepieux@cpt.univ-mrs.fr}
}


%


\maketitle

\begin{abstract}
\boldmath
In this work, we focus on the finite frequency current-current correlations between edge states in a fractional quantum Hall two dimensional gas and on their relations to the quantum admittance. Using a refermionization method, we calculate these quantities within the same framework. Our results apply whatever the values of backscattering amplitude, frequency, voltage and temperature, allowing us to reach different regimes. Auto-correlations and cross-correlations exhibit distinct frequency dependencies that we discuss in detail.

\end{abstract}


%
\IEEEpeerreviewmaketitle

\section{Introduction}
Both the current noise and the admittance provide information about the dynamics of mesoscopic conductors. However, whereas the current fluctuations have been widely studied in interacting systems, it is not the case of the quantum admittance for which only few theoretical works are available \cite{buttiker93,hamamoto10,cottet11,crepieux13,filippone13}. Moreover, the inter-relations between these quantities have not been deeply explored despite the fact that they verify the simple relation: $2\hbar\omega\mathrm{Re}[Y(\omega)]=S(-\omega)-S(\omega)$, which means that a non-vanishing real part of the admittance $Y$ is responsible of the asymmetry of the noise $S$ obtained in interacting systems \cite{safi08,safi11,zamoum12}. The admittance is a quantity which deserves to be known, especially as its experimental determination expends rapidly. Indeed, following the pioneer works by Gabelli and co-workers \cite{gabelli06,gabelli07,gabelli12}, measurements of the quantum admittance in various mesoscopic systems have been achieved recently. In particular, high-frequency admittance has been measured for a quantum point contact in the quantum Hall regime \cite{hashisaka12}, determination of the admittance has been performed for a carbon nanotube double quantum dot \cite{chorley12}, both high-frequency admittance and noise have been measured in superconductor/insulator/superconductor junction \cite{basset12a}, and dynamic admittance has been studied in a quantum dot coupled to a two dimensional electron gas \cite{frey12}. It is thus needed to develop in parallel theoretical studies of quantum admittance. Here, we consider a two dimensional electron gas in the fractional quantum Hall regime and we carefully look at the inter-relations between admittance, which was calculated in Ref. \cite{crepieux13}, and current-current correlations between edge states along which the fractional charges are propagated.

The paper is organized as follows: In Sec. II, we present the model used to describe the chiral edge states in the fractional quantum Hall regime and the method of calculation of the current-current correlations generated by a constriction (see figure \ref{diagrams}). In Sec. III, we give the expressions of the auto- and cross-correlators in terms of current, admittance and integral which involve all the transmission amplitude through the constriction. In Sec. IV, we discuss the results in various limits. We conclude in Sec. V.

\section{Model}

A two-dimensional electron gas in the fractional quantum Hall regime is modeled in the framework of the Tomonaga-Luttinger theory \cite{tomonaga50,luttinger63} by the Hamiltonian:
\begin{eqnarray}\label{hamiltonian}
H&=&\frac{\hbar v_\mathrm{F}}{4\pi}\int_{-\infty}^{\infty}dx\Big[(\partial_x\phi_L(x))^2+(\partial_x\phi_R(x))^2\Big]\nonumber\\
&&+\frac{\Gamma_\mathrm{B}}{4\pi}e^{i[\phi_L(x)+\phi_R(x)]/\sqrt{2}-i\overline{e}\chi(\tau)/(\hbar c)}+hc~,
\end{eqnarray}
where $\phi_{L}$ and $\phi_{R}$ are the bosonic fields associated with the left~($L$) and right~($R$) moving electrons, $v_\mathrm{F}$ is the Fermi velocity, and $\chi(\tau)=-c\int V(\tau)d\tau$ is included in order to treat the applied voltage. The energy $\Gamma_\mathrm{B}$ corresponds to the backscattering amplitude at the constriction (see Fig. \ref{diagrams}) and $\overline{e}=\nu e$ is the fractional charge associated to the filling factor $\nu$. To calculate transport properties, we use a refermionization method \cite{chamon96} which applied for $\nu=1/2$. Note that Eq.~(\ref{hamiltonian}) describes incompressible chiral edge modes which do not correspond to the half filled states since those ones are expected to be compressible \cite{halperin93,murthy03}. However, the results obtained by the refermionization method provide interesting features on the relation between admittance and current correlations in such a system that are worth to study.

\section{Results}

\subsection{Quantum admittance}

The admittance is obtained by calculating the photo-assisted current, i.e., the response to an ac voltage superimposed to a dc voltage: $V(t)=V_\mathrm{dc}+V_\mathrm{ac}\cos(\omega \tau)$, associated to Eq.~(\ref{hamiltonian}) and taking the derivative of its first order harmonic with respect to $V_\mathrm{ac}$ \cite{crepieux13,crepieux04}:
\begin{eqnarray}\label{admittance}
Y(\omega)&=&\frac{e^2}{2h\omega}\int_{-\infty}^{\infty}\Big[t(\Omega)-t(\Omega-\omega)\Big] \nonumber\\
&&\times\Big[f(\hbar\Omega+\overline{e}V_\mathrm{dc})-f(-\hbar\Omega+\overline{e}V_\mathrm{dc})\Big]d\Omega~,
\end{eqnarray}
where $f$ is the Fermi-Dirac distribution function, and $t(\Omega)=(i\Gamma_\mathrm{B}/2)/(\hbar\Omega+i\Gamma_\mathrm{B}/2)$ is the transmission amplitude through the constriction.

\subsection{Current-current correlations}

\begin{figure}[!t]
\centering
\includegraphics[width=8.5cm]{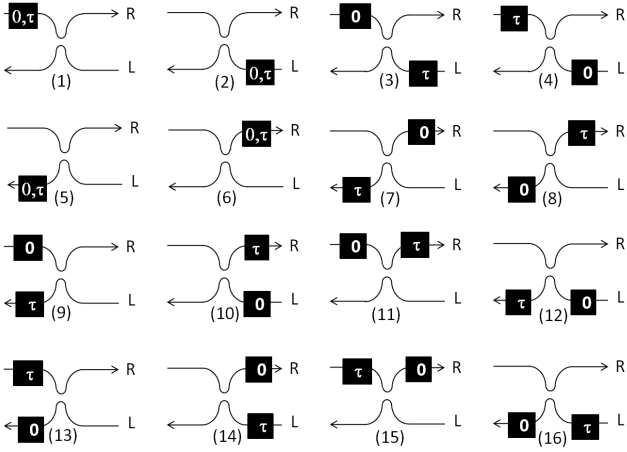}
\caption{Schematic representation of all possible current-current correlators between two chiral edge states coupled by a constriction. The diagrams 1, 2, 5 and 6 gives the auto-correlators, the remaining diagrams correspond to the cross correlators. The values $0$ and $\tau$ in the black boxes indicate the associated time arguments of the current operators in Eq.~(\ref{definition}).}
\label{diagrams}
\end{figure}

Next, we calculate the Fourier transform of the non-symmetrized correlations between currents in branches $i$ and $j$ associated to Eq.~(\ref{hamiltonian}) with $V(t)=V_\mathrm{dc}$:
\begin{eqnarray}\label{definition}
S_{ij}(x,x',\omega)=\int_{-\infty}^\infty dt e^{i\omega t}\langle \delta I_i(x,0)\delta I_j(x',\tau)\rangle~,
\end{eqnarray}
with $\delta I_i(x,\tau)=I_i(x,\tau)-\langle I_i(x,\tau)\rangle$. For chiral edge states, we have $I_i(\tau)=ev_\mathrm{F}r_i\rho_i(x,\tau)$, where $r_i=1$ for right~($i=R$) moving branch and $r_i=-1$ for left~($i=L$) moving branch. $\rho_i(x,\tau)=\psi_i^\dag(x,\tau)\psi_i(x,\tau)$ is the density operator which is related to the density operators $\rho_\pm(x,\tau)$ associated to the new fermionic operators $\psi^\dag_\pm$ and $\psi_\pm$ introduced in the refermionization procedure through the relation \cite{chamon96}: $\rho_i(x,\tau)=[\rho_+(r_ix,\tau)+r_i\rho_-(r_ix,\tau)]/2$. Note that $\psi^\dag_+$ and $\psi_+$ are free fields that are affected neither by the applied voltage nor by the backscattering. Thus, we have:
\begin{eqnarray}
S_{ij}(x,x',\omega)&=&\frac{e^2v_\mathrm{F}^2r_ir_j}{4}\int_{-\infty}^\infty d\tau e^{i\omega \tau}\nonumber\\
&&\times\Big[\langle\delta\rho_+(r_ix,0)\delta\rho_+(r_jx',\tau)\rangle\nonumber\\
&&+r_ir_j\langle\delta\rho_-(r_ix,0)\delta\rho_-(r_jx',\tau)\rangle\Big]~,
\end{eqnarray}
where $\delta \rho_\pm(x,\tau)=\rho_\pm(x,\tau)-\langle \rho_\pm(x,\tau)\rangle$. All the possible correlators found by varying $i$, $j$, $x$ and $x'$ are shown on figure~\ref{diagrams}. Since we have four branches, the total number of correlators equals $2^4$.

\subsubsection{Auto-correlations}

They correspond to $i=j$ and $x=x'$ and take real values. There are four auto-correlators equal in pairs (the indexes used below refer to the diagram numbers in figure~\ref{diagrams}). The calculation gives:
\begin{eqnarray}
S_{1}(\omega)=S_{2}(\omega)=\frac{G_\mathrm{q}\hbar\omega N(\hbar\omega)}{2}~,
\end{eqnarray}
where $G_\mathrm{q}=e^2/\hbar$ is the quantum of conductance, and $N(\hbar\omega)=[\exp(\hbar\omega/(k_\mathrm{B}T))-1]^{-1}$ \cite{note1}. In addition, we have:
\begin{eqnarray}
S_{5}(\omega)=S_{6}(\omega)=S_{0}(\omega)+\frac{G_\mathrm{q}\hbar\omega N(\hbar\omega)}{2}\nonumber\\
-\hbar\omega\big[2N(\hbar\omega)+1\big]\mathrm{Re}\{Y(\omega)\}~,
\end{eqnarray}
where:
\begin{eqnarray}\label{S0_exp}
&&S_{0}(\omega)=\frac{e}{4} \sum_{\sigma=\pm}\sum_{\tilde\sigma=\pm}\Bigg\{
N(\tilde\sigma\hbar\omega+\sigma eV_\mathrm{dc})\nonumber\\
&&\times\Big[I(\tilde\sigma\hbar\omega/\overline{e}+\sigma V_\mathrm{dc})+I(\sigma V_\mathrm{dc})\Big]\nonumber\\
&&-\frac{e}{4\pi}\Big[N(\tilde\sigma\hbar\omega+\sigma eV_\mathrm{dc})-N(\tilde\sigma\hbar\omega)\Big]\nonumber\\
&&\times\int_{-\infty}^{\infty}d\Omega f(\hbar\Omega-\sigma \overline{e}V_\mathrm{dc})\big|t(\Omega+\tilde\sigma\omega)+t^*(\Omega)\big|^2\Bigg\}~,\nonumber\\
\end{eqnarray}
is an even function in terms of frequency and voltage. The dc current reads as:
\begin{eqnarray}\label{current}
&&I(V_\mathrm{dc})=\frac{e}{4\pi}\int_{-\infty}^{\infty}t(\Omega)t^*(\Omega)\nonumber\\
&&\times\Big[f(\hbar\Omega-\overline{e}V_\mathrm{dc})-f(\hbar\Omega+\overline{e}V_\mathrm{dc})\Big]d\Omega~.
\end{eqnarray}
The auto-correlators $S_1$ and $S_2$ are those of charges free to propagate along the edge states (the constriction is not yet reached), whereas the auto-correlators $S_5$ and $S_6$ are those of charges that have been reflected or transmitted through the constriction. It explains why the transmission amplitude $t$ appears in these two last auto-correlators only.

\subsubsection{Cross-correlations}

They corresponds to $i\ne j$ or/and $x\ne x'$ (in the latter case, we take the limits $x$ and $x'$ close to zero in order to focus on what happens near the constriction). There are twelve cross-correlators, some of them take real values whereas the others take complex values. The two last correlators of the first line in figure~1 are equal to zero: $S_{3}(\omega)=S_{4}(\omega)=0$, since as long as the constriction is not reached, the carriers in two independent chiral branches are not correlated. The two last correlators of the second line in figure~\ref{diagrams} are equal to each other: $S_{7}(\omega)=S_{8}(\omega)$, with:
\begin{eqnarray}
S_{7}(\omega)=S_{0}(\omega)-\hbar\omega\big[2N(\hbar\omega)+1\big]\mathrm{Re}\{Y(\omega)\}~.
\end{eqnarray}
The two first correlators of the third and fourth lines in figure~\ref{diagrams} are related through the relation: $S_{9}(\omega)=S_{10}(\omega)=S^*_{13}(\omega)=S^*_{14}(\omega)$, with:
\begin{eqnarray}
S_{9}(\omega)=\hbar\omega N(\hbar\omega)\left[Y(\omega)-\frac{G_\mathrm{q}}{4}\right]~.
\end{eqnarray}
The two last correlators of the third and fourth lines in figure~\ref{diagrams} are related through $S_{11}(\omega)=S_{12}(\omega)=S^*_{15}(\omega)=S^*_{16}(\omega)$ with:
\begin{eqnarray}
S_{11}(\omega)=\hbar\omega N(\hbar\omega)\left[Y(\omega)+\frac{G_\mathrm{q}}{4}\right]~.
\end{eqnarray}
Note that the knowledge of $S_0$ and $Y$ fully determine the auto- and cross-correlators.

\section{Discussion}

In this section, we consider the excess correlators \cite{note2}, i.e. the difference between their values at finite dc voltage $V_\mathrm{dc}$ and their values at zero voltage: $\Delta S_n(\omega)=S_n(\omega)-S_n(\omega)|_{V_\mathrm{dc}=0}$ with $n\in[1,2^4]$. The calculations show that the excess correlators are equal four to four. Since the carriers that are ahead the constriction do not feel the voltage, the excess correlators for the diagrams 1, 2, 3 and 4 of the first line in figure~\ref{diagrams} vanish:
\begin{eqnarray}
\Delta S_{n\in[1,4]}(\omega)=0~.
\end{eqnarray}
The excess correlators associated to diagrams 5, 6, 7 et 8 coincide with each other and are related to $S_0$ and to the real part of the excess admittance:
\begin{eqnarray}\label{deltaS5}
\Delta S_{n\in[5,8]}(\omega)=\Delta S_{0}(\omega)-\hbar\omega\big[2N(\hbar\omega)+1\big]\mathrm{Re}\{\Delta Y(\omega)\}~.\nonumber\\
\end{eqnarray}
The excess correlators associated to diagrams 9, 10, 11 and 12 are related to the excess admittance:
\begin{eqnarray}
\Delta S_{n\in[9,12]}(\omega)=\hbar\omega N(\hbar\omega)\Delta Y(\omega)~,
\end{eqnarray}
as well as the diagrams 13, 14, 15 and 16:
\begin{eqnarray}
\Delta S_{n\in[13,16]}(\omega)=\hbar\omega N(\hbar\omega)\Delta Y^*(\omega)~.
\end{eqnarray}
The expressions of the total excess noise $\Delta S_T$ and backscattering excess noise $\Delta S_\mathrm{B}$ can also be derived. Indeed from figure~\ref{diagrams}, we deduce from the definitions of the correlators that $S_\mathrm{T}=S_1+S_5+S_9+S_{13}$ and $S_\mathrm{B}=S_1+S_6-S_{11}-S_{15}$, which lead to:
\begin{eqnarray}\label{noise_T}
\Delta S_\mathrm{T}(\omega)&=&\Delta S_0(\omega)-\hbar\omega\mathrm{Re}\{\Delta Y(\omega)\}~,\\
\Delta S_\mathrm{B}(\omega)&=&\Delta S_T(\omega)-4\hbar\omega N(\hbar\omega)\mathrm{Re}\{\Delta Y(\omega)\}~.
\end{eqnarray}
Since $\Delta S_0$ is even in frequency and the real part of $\Delta Y$ is odd in frequency, Eq.~(\ref{noise_T}) verifies the relation $\Delta S_\mathrm{T}(-\omega)-\Delta S_\mathrm{T}(\omega)=2\hbar\omega \mathrm{Re}\{\Delta Y(\omega)\}$ as it should. In the following we study the frequency profile of the excess correlators with emphasize on various limits: first the low temperature limit in order to understand how the backscattering amplitude strength affects this profile, next the weak backscattering regime in order to characterize the effect of temperature, and finally the high temperature limit.

\subsection{Low temperature limit}

\begin{figure}[!h]
\centering
\includegraphics[width=6.2cm]{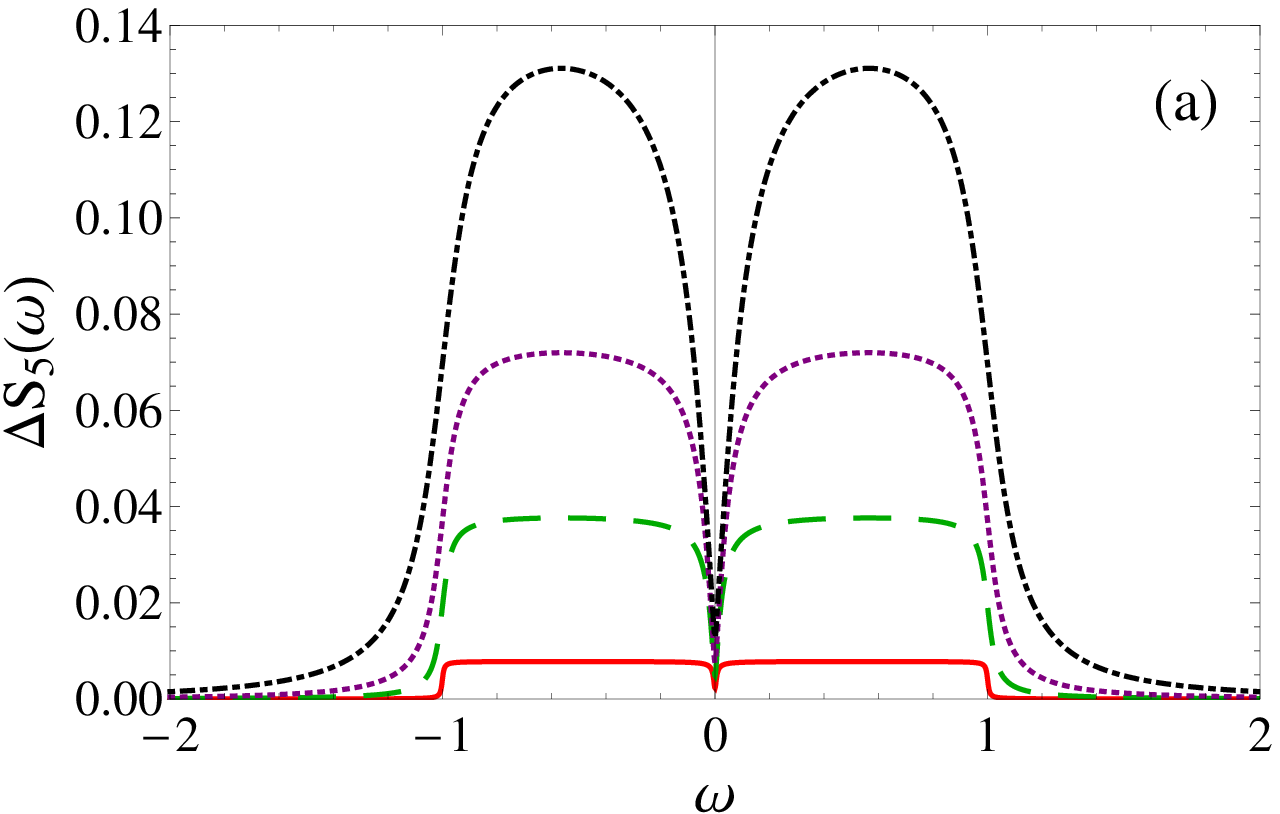}
\includegraphics[width=6.2cm]{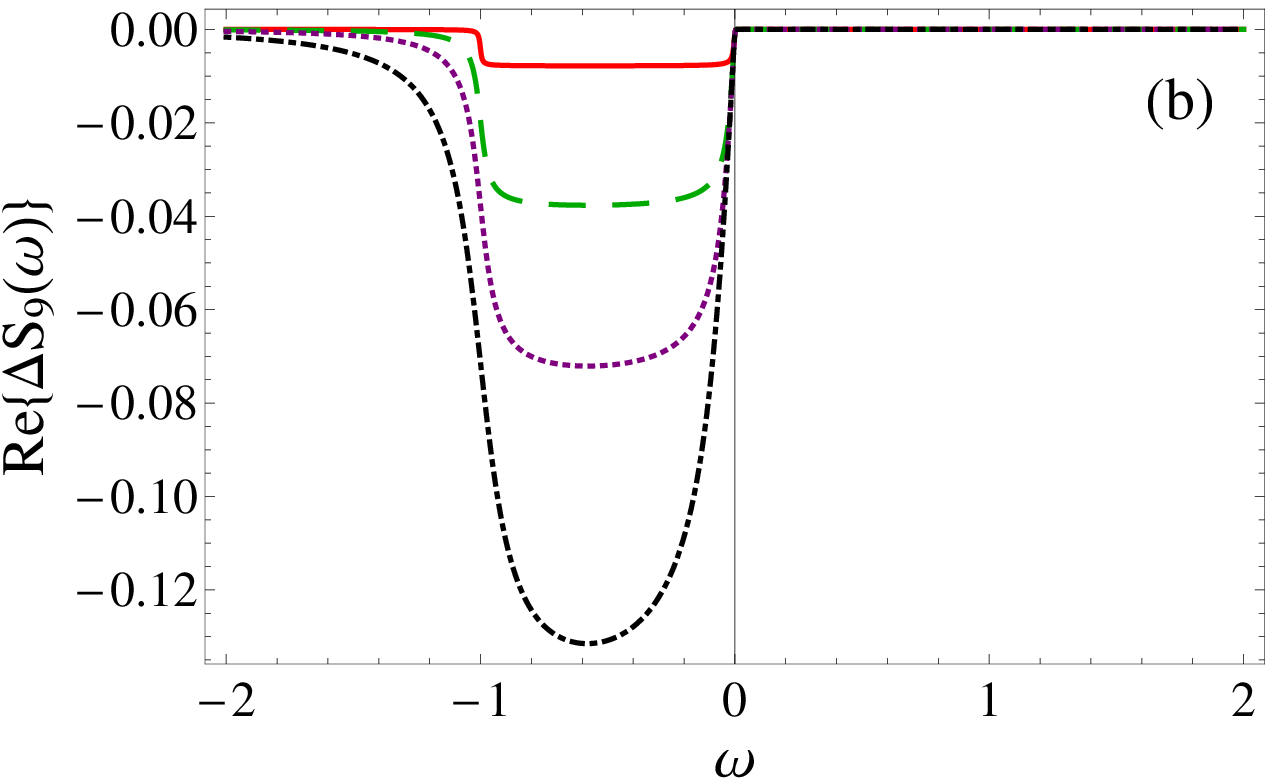}
\includegraphics[width=6.2cm]{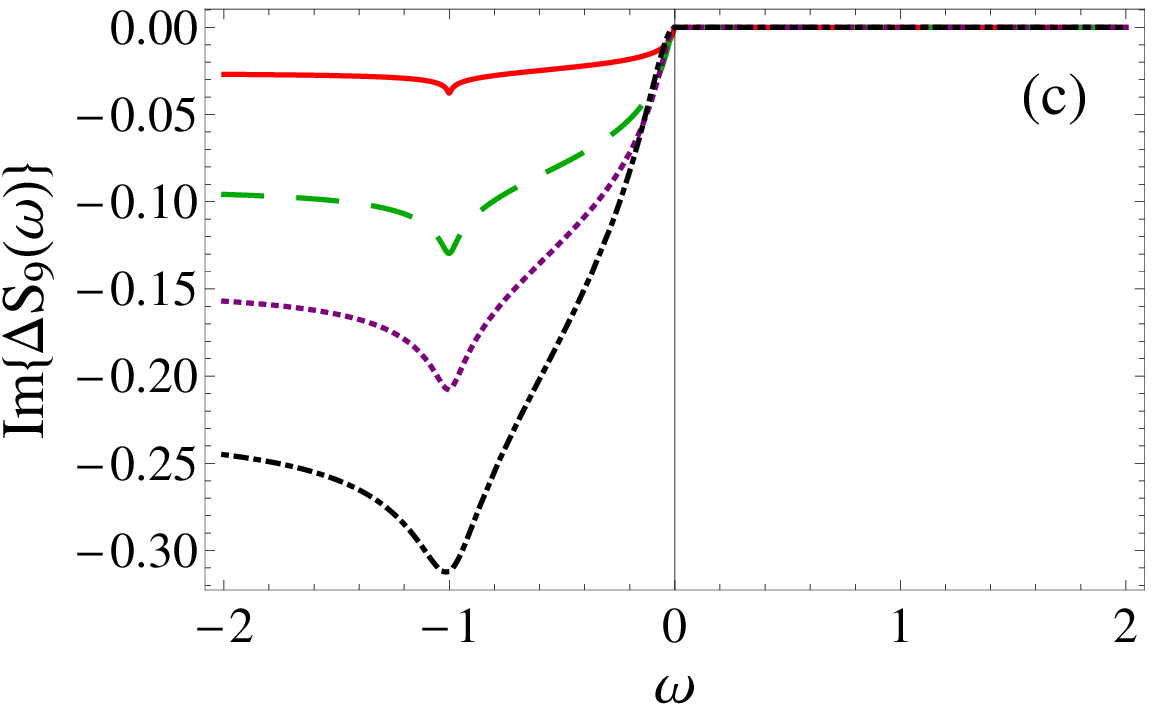}
\caption{(a) Excess correlator $\Delta S_5$, (b) real part and (c) imaginary part of $\Delta S_9$ in units of $\bar{e}V_\mathrm{dc}G_\mathrm{q}$, as a function of frequency in units of $\bar{e}V_\mathrm{dc}/\hbar$, for various values of the backscattering amplitude $\Gamma_\mathrm{B}$ at temperature $k_\mathrm{B}T=0.001\bar{e}V_\mathrm{dc}$.  The backscattering amplitudes are $\Gamma_\mathrm{B}=0.01\bar{e}V_\mathrm{dc}$ (solid red line), $\Gamma_\mathrm{B}=0.05\bar{e}V_\mathrm{dc}$ (dashed green line), $\Gamma_\mathrm{B}=0.1\bar{e}V_\mathrm{dc}$ (purple dotted line) and $\Gamma_\mathrm{B}=0.2\bar{e}V_\mathrm{dc}$ (dash-dotted black line).}
\label{figure2}
\end{figure}

In figure \ref{figure2}(a) is plotted the correlator $\Delta S_{5}$ at temperature much smaller than the applied voltage. We observe that this excess correlator is symmetric in frequency. It is due to the fact that the diagrams 5, 6, 7 and 8 of figure \ref{diagrams} corresponding to this correlator are symmetrical under time inversion, thus $\langle \delta I_i(x,0)\delta I_j(x',\tau)\rangle=\langle \delta I_i(x,\tau)\delta I_j(x',0)\rangle$ which immediately leads to $\Delta S_{5}(\omega)=\Delta S_{5}(-\omega)$. The real and imaginary parts of $\Delta S_{9}$ are plotted in figures \ref{figure2}(b) and \ref{figure2}(c). This excess correlator has the very particular property to be non zero only at negative frequency. We can interpret this result as a time inversion symmetry breaking which is understandable by looking at diagrams 9, 10, 11 and 12 of figure \ref{diagrams}. Comparing now figures \ref{figure2}(a) and \ref{figure2}(b), we remark that  at negative frequency we have $\Delta S_5(\omega)\approx-\mathrm{Re}\{\Delta S_9(\omega)\}$, which means that the contribution of the term $\Delta S_0$ in Eq. (\ref{deltaS5}) is negligible in that regime. Since at zero temperature, we have $N(\hbar\omega)=-\Theta(-\omega)$, where $\Theta$ is the Heaviside function, we can approximate the correlators as:
\begin{eqnarray}
\Delta S_{5}(\omega)&\approx&-\hbar|\omega|\mathrm{Re}\{\Delta Y(\omega)\}~,\\
\Delta S_{9}(\omega)&\approx&-\hbar\omega\Theta(-\omega)\Delta Y(\omega)~,\\
\Delta S_\mathrm{T}(\omega)&\approx&-\hbar\omega\mathrm{Re}\{\Delta Y(\omega)\}~,\\
\Delta S_\mathrm{B}(\omega)&\approx&\hbar\omega\big[4\Theta(-\omega)-1\big]\mathrm{Re}\{\Delta Y(\omega)\}~.
\end{eqnarray}
Thus, from the fact that $\Delta S_{0}(\omega)\approx 0$ in the low temperature limit, we find that the excess correlators are entirely determined by the excess admittance in that regime. As a consequence, the singularities observed at $\hbar\omega=\pm\bar{e}V_\mathrm{dc}$ in figures \ref{figure2}(a), \ref{figure2}(b) and \ref{figure2}(c) for weak backscattering amplitudes are those of the admittance (see figure 2 of Ref. \cite{crepieux13}). These singularities disappear when the backscattering amplitude increases. Note that since the correlators $\Delta S_5$ and $\Delta S_6$ appearing in the definitions of the total and backscattering excess noises are symmetric in frequency, the asymmetry of $\Delta S_T$ and $\Delta S_B$ comes from the diagrams 9, 11, 13 and 15 of figure \ref{diagrams}.

\subsection{Weak backscattering limit}

\begin{figure}[!h]
\centering
\includegraphics[width=6.2cm]{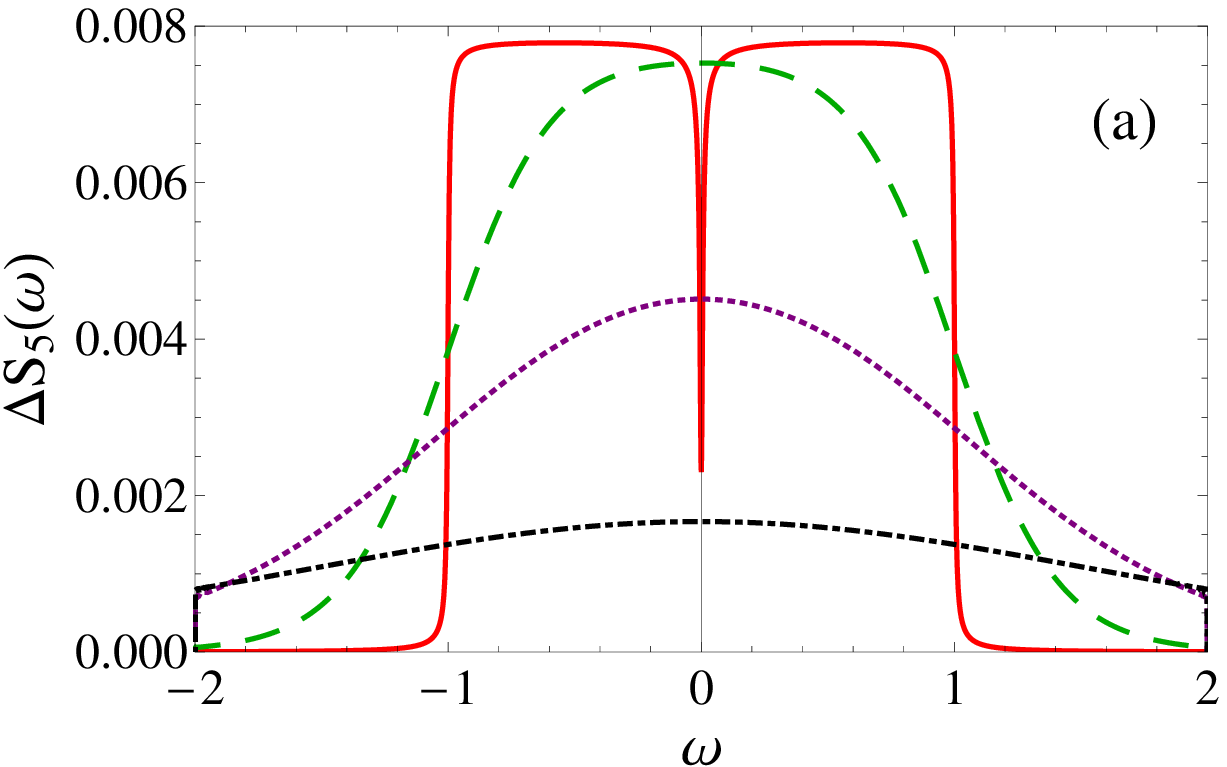}
\includegraphics[width=6.2cm]{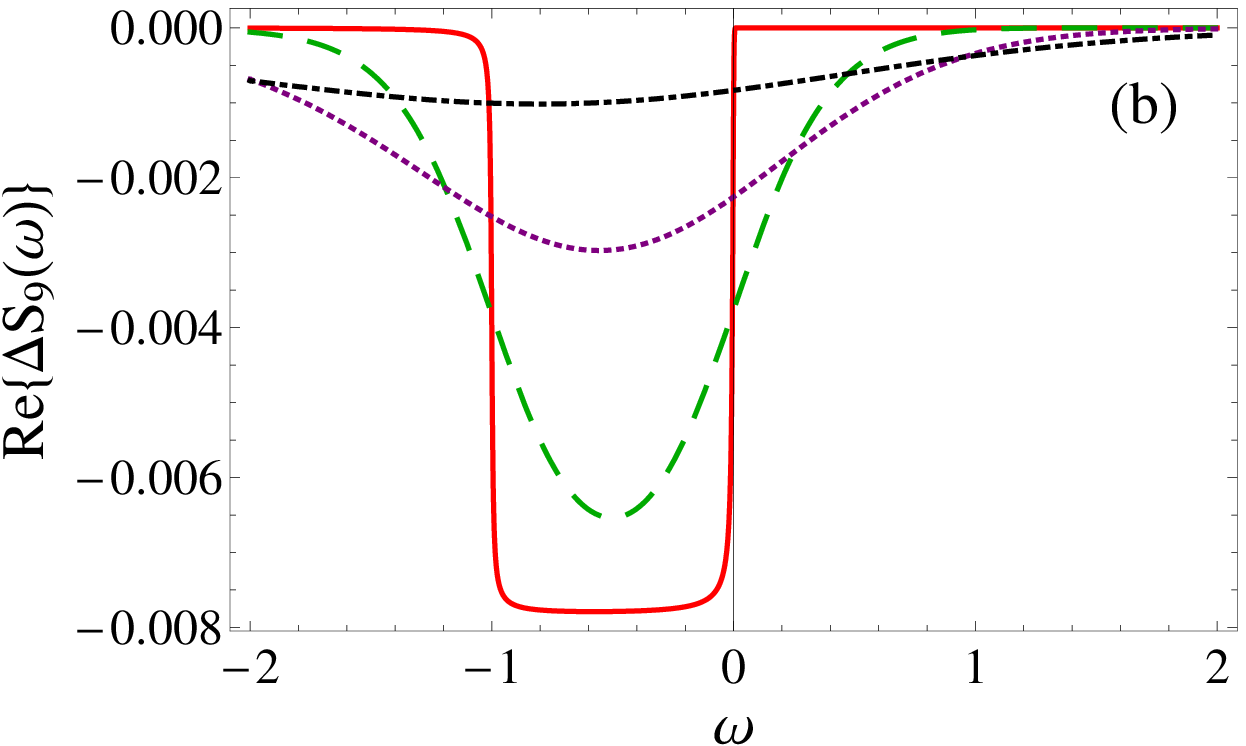}
\includegraphics[width=6.2cm]{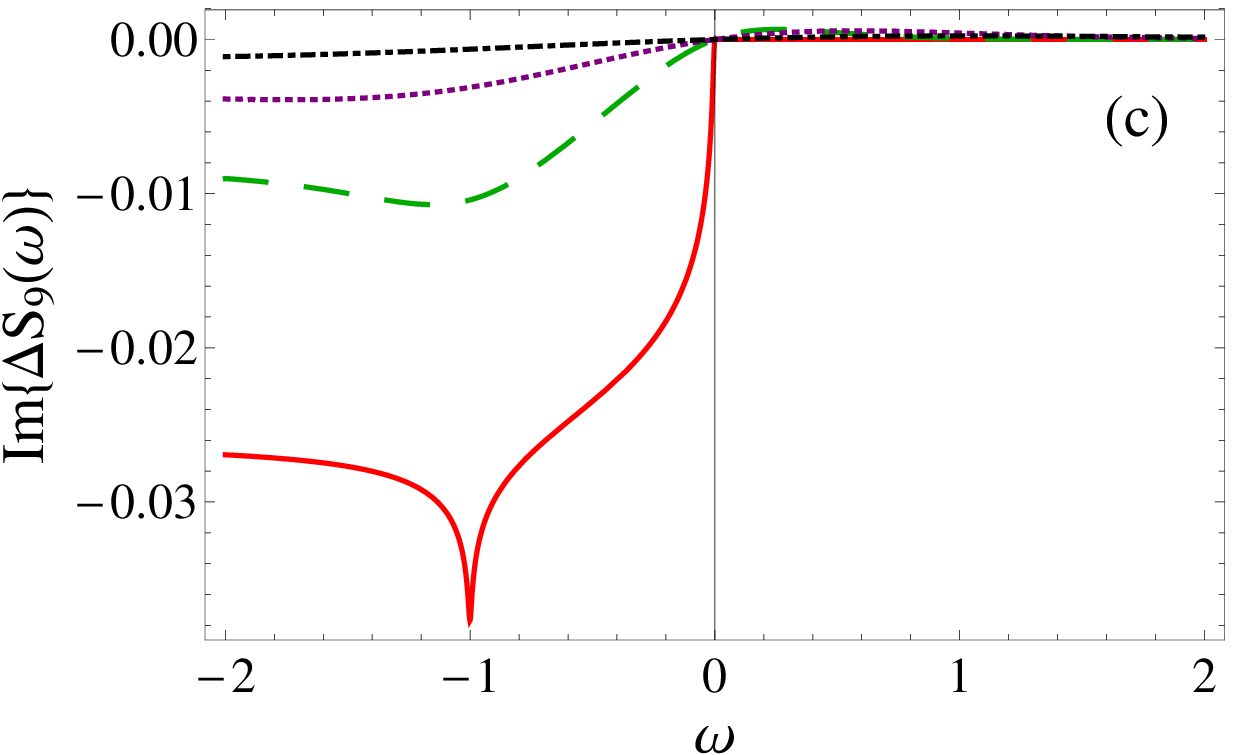}
\caption{(a) $\Delta S_5$ in units of $\bar{e}V_\mathrm{dc}G_\mathrm{q}$, as a function of frequency in units of $\bar{e}V_\mathrm{dc}/\hbar$, for various temperature at weak backscattering amplitude $\Gamma_\mathrm{B}=0.01\bar{e}V_\mathrm{dc}$. The temperatures are $k_\mathrm{B}T=0.001\bar{e}V_\mathrm{dc}$ (solid red line), $k_\mathrm{B}T=0.2\bar{e}V_\mathrm{dc}$ (dashed green line), $k_\mathrm{B}T=0.5\bar{e}V_\mathrm{dc}$ (purple dotted line) and $k_\mathrm{B}T=\bar{e}V_\mathrm{dc}$ (dash-dotted black line). The same curves for (b) the real part and (c) the imaginary part of $\Delta S_9$.}
\label{figure3}
\end{figure}

We turn now our attention to the behavior of the correlators when the backscattering amplitude is much smaller than the voltage. In figures \ref{figure3}(a), \ref{figure3}(b) and \ref{figure3}(c) are plotted $\Delta S_{5}$ and the real and imaginary parts of $\Delta S_9$ at various temperature for $\Gamma_\mathrm{B}=0.01\bar{e}V_\mathrm{dc}$. The correlator $\Delta S_{5}$ is equal to zero at $|\hbar\omega|>\bar{e}V_\mathrm{dc}$ under the condition that both temperature and backscattering are small in comparison to the voltage (see solid red lines on figures \ref{figure2}(a) and \ref{figure3}(a)). Concerning the correlator $\Delta S_{9}$, both the backscattering and the thermal effects are able to give a non-vanishing contribution at $\hbar\omega<-\bar{e}V_\mathrm{dc}$, whereas only thermal effects can give a non-vanishing contribution at $\hbar\omega>0$, which stays however small for what concerns the imaginary part of $\Delta S_9$ (see figure \ref{figure3}(c)).

\subsection{High temperature limit} 

For temperature much higher than the voltage, i.e., at equilibrium, we find from Eq. (\ref{S0_exp}) that:
\begin{eqnarray}
S_0(\omega)\approx \hbar\omega\big[2N(\hbar\omega)+1\big]\mathrm{Re}\{Y(\omega)\}~,
\end{eqnarray}
which leads to the cancellation of $\Delta S_5$ in agreement with the evolution of the curves shown in figure \ref{figure3}(a), and to:
\begin{eqnarray}
\Delta S_\mathrm{T}(\omega)\approx
-\Delta S_\mathrm{B}(\omega)\approx 2\hbar\omega N(\hbar\omega)\mathrm{Re}\{\Delta Y(\omega)\}~,
\end{eqnarray}
in agreement with the fluctuation-dissipation theorem since the real part of the admittance correspond to the ac conductance. The cancellation of $\Delta S_5$ is related to the fact that the dominant contribution in that limit is the thermal noise which is voltage independent. Since the excess correlator is defined as the difference between its values at finite and zero voltages, there is an exact compensation. From these results, we understand that in that regime too, the excess correlators are entirely determined by the excess admittance.

\section{Conclusion}
Our study shows that the current, the quantum admittance and  the current-current correlators between the edge states of a fractional quantum Hall gas are quantities that are closely related. Indeed, since all these quantities can be expressed in terms of the transmission amplitude through the constriction which coupled the edge states, it is possible to obtain direct relations between them. Accordingly, the singularities that are observed in the correlators in the weak backscattering and low temperature regimes at frequencies equal to $\pm \bar{e}V_\mathrm{dc}/\hbar$, are directly related to the singularities of the quantum admittance. Moreover, in both low and high temperature limits, we have shown that all the current correlators are fully determined by the admittance and we have established explicit relations between them. These relations can be use to find the value of the finite-frequency noise from the measurement of the admittance which is easier to perform.


\section*{Acknowledgment}

The authors would like to thank T. Martin and I. Safi.




\begin{thebibliography}{99}
\bibitem{note1} Note that at positive frequency $N(\hbar\omega)$ corresponds to the Bose-Einstein distribution function.
\end{thebibliography}
%

\end{document}